# Advanced DFT-NEGF transport techniques for novel 2D-material and device exploration including HfS$_2$/WSe$_2$ van-der-Waals Heterojunction TFET and WTe$_2$/WS$_2$ metal/semiconductor contact.

A. Afzalian[1], E. Akhoundi[1,2], G. Gaddemane[1,2], R. Duflou[1,2] and M. Houssa[1,2]

*Abstract*— We present, here, advanced DFT-NEGF techniques that we have implemented in our ATOmistic MOdelling Solver, ATOMOS, to explore transport in novel materials and devices and in particular in van-der-Waals heterojunction transistors. We describe our methodologies using plane-wave DFT, followed by a Wannierization step, and linear combination of atomic orbital DFT, that leads to an orthogonal and non-orthogonal NEGF model, respectively. We then describe in detail our non-orthogonal NEGF implementation including the Sancho-Rubio and electron-phonon scattering within a non-orthogonal framework. We also present our methodology to extract electron-phonon coupling from first principle and include them in our transport simulations. Finally, we apply our methods towards the exploration of novel 2D materials and devices. This includes 2D material selection and the Dynamically-Doped FET for ultimately scaled MOSFETS, the exploration of vdW TFETs, in particular the HfS$_2$/WSe$_2$ TFET that could achieve high on-current levels, and the study of Schottky-barrier height and transport through a metal-semiconducting WTe$_2$/WS$_2$ VDW junction transistor.

*Index Terms*— Density functional theory, Heterojunctions, MOSFETs, Semiconductor device modeling, Tunnel transistors, Two dimensional material, Quantum effect semiconductor devices, Quantum theory.

## I. Introduction

Two dimensional materials and in particular transition metal dichalcogenides (TMDs) are widely investigated by the scientific community nowadays [1-6]. Their large variety of bandgaps, effective masses, and their excellent electrostatic properties related to their 2D nature hold promise to find the candidate for ultra-scaled CMOS applications. In addition, 2D van der Waals (vdW) heterojunction could be a powerful tool to enlarge their application space, such as in band-to-band tunneling field-effect-transistors (TFETs) for low-power electronics [1-4], or 2D metal-2D semiconductor for low Schottky barrier contacts [5].

Non-ab-initio models fitted on density functional theory (DFT) band structures are widely used for predicting performances of 2D material transistors [3,6,7,8], due to their wider availability and strongly reduced computational load. However, such models tend to predict drive current larger than those obtained by the DFT models [6]. To accurately model 2D-material intricate band structures and transport effects, as well as to accurately capture the vdW interlayer coupling coefficient on which the tunneling current is highly sensitive [3], a DFT-based quantum transport method, such as DFT-NEGF (non-equilibrium Green's functions), is ideal [2,6]. Different challenges and choices arise, however, for the modeling of these low-dimensional materials using DFT methods and some will be discussed here.

The choice of the DFT model that is to be used as the building block to get the require material properties, i.e., the basic Hamiltonian (*H*) elements from which the full device *H* is built, will be discussed in section II.a. Plane-wave methods, such as the projector augmented wave (PAW)-DFT, are popular and softwares like VASP [9] or QUANTUM ESPRESSO (QE) [10] are widely used. For transport, however, a localized basis, typically a linear combination of atomic orbitals (LCAO), is preferred for efficiency [11]. A popular method is to convert from the PAW basis to an orthogonal LCAO basis using the Wannierization technique [12]. A full description of such a method and its implementation in our ATOmistic MOdelling Solver (ATOMOS) is described in [6, 13] for homojunction materials. Here, we will detail the procedure we used for the case of a vdW heterojunction. The modelling of heterojunctions or materials with defects may, however, require large supercells that are more efficiently handled directly using non-orthogonal

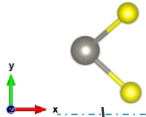
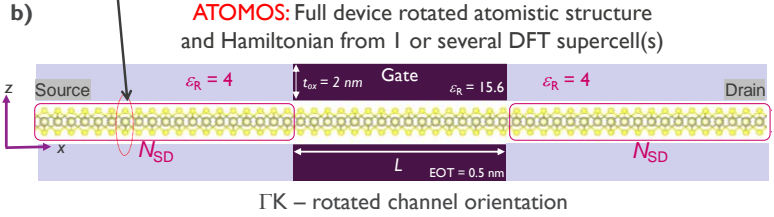
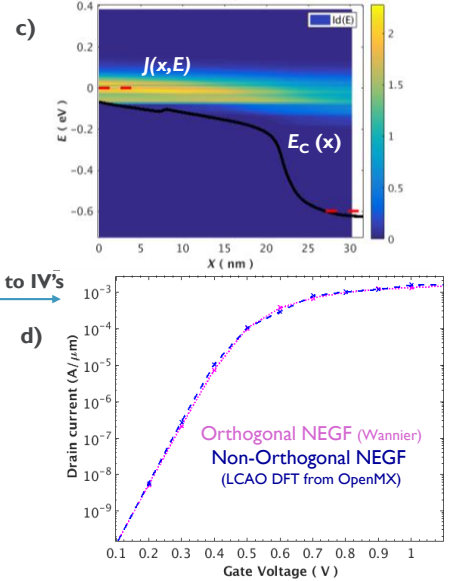

Fig. 1. The DFT-NEGF simulation flow for material/device exploration in ATOMOS is illustrated for a 2D double-gated DG $L$ = 10 nm $WS_2$ transistor in ATOMOS using PBE: a) DFT simulation to get material properties and $H$ building blocks, b) import and build the full device geometry and $H$ in ATOMOS. Perform DFT-NEGF simulations to get c) spectrally resolved internal quantities such as current spectrum $J(x,E)$ and d) external quantities such as $I_D(V_G)$ characterisitics. In d) the curves including e-ph for both the orthogonal (PAW DFT using QE+Wannier) and non-othogonal (LCAO DFT from OPENMX) NEGF models are shown.

LCAO – DFT techniques and codes like OPENMX [14] or CP2K [15]. In addition, using LCAO DFT suppresses the need for the Wannierization process that is serial and becomes prohibitively long for supercells with a large number of atoms, albeit introducing the complication of a non-orthogonal NEGF formalism. We have implemented such a formalism in our simulator and its implementation will be discussed in section II.B.

Ballistic transport is not sufficient to properly describe transport properties of many 2D materials [6]. The electron-phonon (e-ph) coupling is usually not known for novel materials and needs to be computed. Our methodology to compute the e-ph matrix using first-principle methods [16] and apply it to our transport simulations will be discussed in section II.C.

Finally, in a last part, we will showcase the potential of our simulator and methods towards the exploration of 2D materials and devices. In section III.A, we will focus on 2D material selection and the Dynamically-Doped FET ($D_2$-FET) [6] for ultimately scaled CMOS. In section III.B, we will focus on the physics, design and performance of the vdW TFETs using a Wannierized $H$. In particular, our results predict that the $HfS_2/WSe_2$ TFET could achieve on-current levels as high as 420 µA/µm at a supply voltage of $V_{DD}$ = 0.35 V and $I_{OFF}$ = 100 pA/µm. This is, to the best of our knowledge, by far, the highest current reported in a vdW TFET using accurate DFT-NEGF simulations [2]. On the other hands, our results predict that the $MoS_2/WSe_2$ TFET only achieve a disappointing drive current of 14 µA/µm, which is about 2 order of magnitude lower than what was predicted using non-ab-initio methods [7,8]. Finally, in section III.C, we will apply the ability to handle a large supercell using LCAO DFT and our non-orthogonal NEGF model to study the Schottky-barrier height ($SB_H$) and transport through a metal-semiconducting $WTe_2/WS_2$ vdW-junction transistor.

## II. METHOD

Our quantum-transport solver, ATOMOS [6], was specifically developed for high-performance computing and the use of computationally-heavy DFT Hamiltonians. It is written in C++ and uses multi-threaded MPI with various levels of parallelism. Any heavy vector-matrix or matrix-matrix operations are performed using BLAS and LAPACK. ATOMOS core transport solver is a dissipative NEGF solver based on the recursive Green's function (RGF) algorithm [6]. To ensure efficient load-balancing, a dynamic scheduler is used to distribute the various energy-momentum ($e$-$k$) points between the different parallel ranks. For optimally generating the energy points, we rely on a recursive adaptive-grid algorithm. For self-consistency, a parallel Poisson solver is used. To expedite the Poisson-NEGF convergence, we employ a predictor-corrector method using the Newton scheme [6]. The anisotropic dielectric permittivities are taken from [17].

### A. DFT Hamiltonian and representation method

As described in Fig. 1a, the first step towards transport simulations of a given material is a first-principle geometry relaxation of its primitive unit cell, followed by an electronic-structure calculation. We support PAW-DFT, followed by a Wannierization step that leads to an orthogonal basis set [6]. We used the DFT package QE and the generalized gradient approximation with the PBE or optB86b exchange-correlation functionals [18]. These functionals typically predict TMD band gaps in close agreement with their experimental values [2,6]. To model the electronic states in the HfS$_2$/WSe$_2$ heterostructure shown in Fig. 2, Van der Waals interactions are included through the DFT-D3 method of Grimme [19]. A plane-wave cutoff energy of 1088 eV (80 Ry) and an 8 × 8 × 1 Monkhorst-Pack *k*-point grid was used in the electronic band structure calculations without spin-orbit coupling. It was verified that the total energy was well converged for these values. The convergence criteria are set to less than 10$^{-3}$ eV/Å forces acting on each ion and a total energy difference smaller than 10$^{-3}$ eV between two subsequent iterations. To cut off the periodic images along the z-direction (Fig. 2a), a vacuum layer of 25 Å was employed in the DFT simulations.

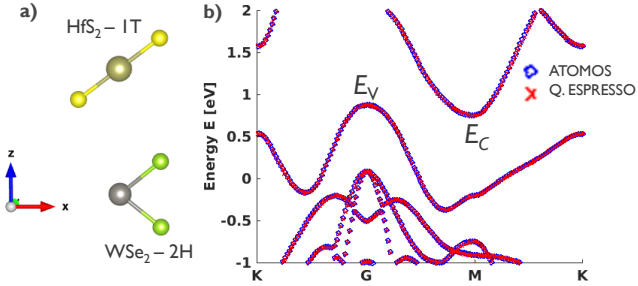

Fig. 2. a) HfS$_2$/WSe$_2$ heterostack relaxed geometry and b) DFT-computed band structure (QUANTUM ESSPRESSO, optB86b) and with ATOMOS using the Wannierized Hamiltonian.

First the equilibrium lattice constants of HfS$_2$ – 1T and WSe$_2$ – 2H monolayers are determined, 3.61 Å and 3.32 Å respectively. The 6-atom-only heterostructure supercell is constructed after applying 4% strain on both materials so that they share a common lattice parameter of $a_0$ = 3.47 Å. The ions are then relaxed again within the stacked unit cells. The Bloch wavefunctions are then transformed into maximally-localized Wannier functions (MLWF) typically centered on the ions using the wannier90 package [20]. Figure 2b demonstrates the validity of our MLWF representation.

ATOMOS also directly supports LCAO DFT with a non-orthogonal model that will be described below. The procedure is similar to that described for PAW DFT but does not require a Wannierization step (that becomes too slow for larger supercells). We used OPENMX to model the WTe$_2$/WS$_2$ heterojunction and the generalized gradient approximation with the PBE exchange-correlation functionals and the DFT-D3 method of Grimme. After relaxation, the in-plane equilibrium lattice vectors of the orthorhombic WTe$_2$ – 1T' supercell with 6 atoms are [3.46777, 0, 0] and [0, 6.255376, 0] Å, while that of the hexagonal WS$_2$ – 2H with 3 atoms are [2.72293, -1.574726, 0] and [2.72293, 1.574726, 0] Å. A 6 atoms orthorhombic supercell is then created for WS$_2$ using the [1,1,0] and [-1,1,0] miller indices. The 12-atom heterostructure supercell directly resulting from straining both materials to a common lattice leads to a 9% residual strain in the y-direction. In turn, this yields a large 1 eV drop in the bandgap of WS$_2$. By tripling the WTe$_2$ cell in the x-direction and doubling the orthorhombic WS$_2$ cell in both the x and y directions, a 42 atoms orthorhombic supercell with a small residual strain of 2.4% and 0.3% in the x and y direction respectively is achieved. For the final supercell, a cutoff energy of 300 Ry and a 7 × 7 × 1 Monkhorst-Pack *k*-point grid was used. For S and Te, a set of

8 pseudo atomic orbitals (PAO) per atoms (1s, 3p and 5d) were used with a cutoff radius of 4.75 Å, while for W, a set of 12 PAOs (2s, 6p and 5d) with a cutoff of 3.7 Å were used.

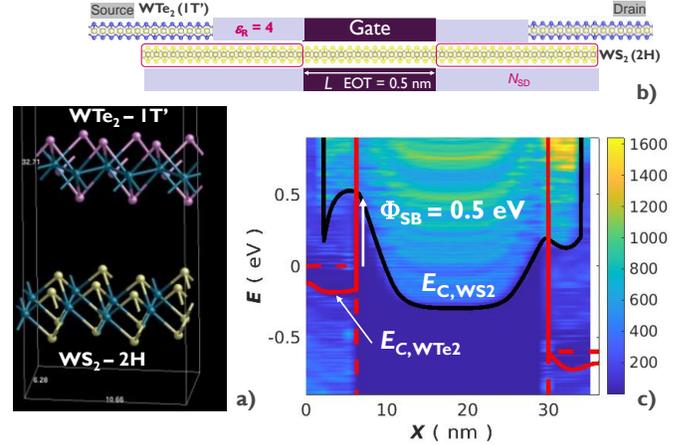

Fig. 3. a) WTe$_2$/WS$_2$ heterostack relaxed geometry and b) Schematic of the simulated *L* = 14 nm double-gated (DG) MOSFET. c) DFT-NEGF computed DoS*(E, x)* (surface plot), with ATOMOS that shows the 0.5 eV Schottky barrier created at the mettalic WTe$_2$ - semiconducting WS$_2$ interface. The red line is the conduction band edge in the top layer (WTe$_2$ and oxide), while the black line is the conduction band edge in the bottom layer along the channel direction in on-state.

As described in Fig. 1b, ATOMOS uses the resulting supercell information, i.e., atom positions, lattice vectors, and the localized MLWF or PAO generated Hamiltonian matrix elements, as building blocks to create the full-device atomic structure and Hamiltonian matrix. We kept in the device Hamiltonian, the required Hamiltonian longer-range interactions (typically 12 to 15 Å for Wannier and about 5 Å for the LCAO case). ATOMOS can further rotate the device geometry to a preferential channel orientation within the 2D layer. We assumed periodic boundary conditions in the width (y-axis) direction. They were modeled with 12 $k_Y$ points on half the Brillion zone, the other half is obtained by symmetry. Transport calculations are then performed using our self-consistent real-space NEGF solver. As a result, spectrally resolved quantities such as current spectrum (Fig. 1c), DoS (Fig. 3c), or occupied and empty states ($G^{<,>}$) or carrier density and current (Fig. 1d) are available.

### B. *Non-orthogonal NEGF model*

OpenMX and CP2K LCAO based DFT tools can be used to generate *H* and the overlap matrix, *S*, to be loaded into ATOMOS. Both packages are in a non-orthogonal basis representation. As a result, the overlap matrix *S* is no longer an identity matrix. It has the same tri-diagonal block sparse structure as *H*. The equations for the retarded ($G^R$), lesser ($G^<$) and greater ($G^>$) Green's functions read [22]:

$$G^R = (ES - H - \Sigma^R)^{-1}, \qquad (1)$$
$$G^< = G^R \Sigma^< G^{R\dagger}, \qquad (2)$$
$$G^> = G^R - G^{R\dagger} + G^<. \qquad (3)$$

*E* is the scalar energy. *S*, *H*, and $\Sigma^{R,<}$ the retarded, lesser self-energies that include the interaction terms (e.g., with the semi-infinite leads $\Sigma_C^{R,<}$ and the electron – phonon scattering terms $\Sigma_S^{R,<}$) are matrices of rank $N$, the total number of atoms in the device × the number of orbitals/atoms. We efficiently store $H$, $S$ and other $G$ matrices using our dedicated sparse block-matrix class, that we specifically customized for the RGF method.

By taking the diagonal elements of the density matrix $\rho$, the electron density $n$ can be calculated

$$n = \int \frac{dE}{2\pi} [\rho]_{diag} \, ; \, \rho = -iG^< S \qquad (4)$$

This density is fed to our parallel Poisson solver. At iteration $i$, after the new electrostatic potential energy $V_i$ has been computed, the potential energy variation is added to $H$ in a symmetric fashion as follows [22]:

$$\left(H_i = H_{i-1} + \tfrac{1}{2}(SdV + dVS)\right) \qquad dV = V_i - V_{i-1} \qquad (5)$$

Electron-phonon scattering is considered using the self-consistent Born approximation [21]. Assuming the phonons stay in equilibrium, and that the e-ph coupling matrix $M_q$ is expressed in an orthogonal basis (e.g., by a deformation potential), the orthogonal scattering self-energy may be written as [6,11,21]:

$$\Sigma_{S,\perp}^<(\mathbf{r}_i, \mathbf{r}_j, E) = \int \frac{d\mathbf{q}}{(2\pi)^3} e^{i\mathbf{q}\cdot(\mathbf{r}_i-\mathbf{r}_j)} |M_q|^2 \times \left(N_q + \tfrac{1}{2} \pm \tfrac{1}{2}\right) G^<(\mathbf{r}_i, \mathbf{r}_j, E \pm \hbar\omega_q).S \qquad (6)$$

where $\mathbf{q}$ and $\omega_q$ are the phonon wave vector and the corresponding angular frequency, $\hbar$ is the reduced Plank's constant, $N_q$ is the phonon-occupation number. The non-orthogonal scattering self-energy is then obtained as:

$$\Sigma_S^< = \tfrac{1}{2}\left(S\Sigma_{S,\perp}^< + \Sigma_{S,\perp}^< S\right) \qquad (7)$$

The contact self-energies are computed with the Sancho-Rubio method [23]. The method proposed in Sancho *et. al* can be extended for a non-orthogonal basis [24]. We assume a homogeneous semi-infinite lead. By definition we have $(\omega S - H)G(\omega) = I$ where $\omega = E + i\eta$. $G(\omega)$ and $G_{00}$ are the lead- and surface-Green's functions, respectively. By taking the matrix elements of $(\omega S - H)G(\omega) = I$ and following the steps detailed by Sancho *et. al* [23] we get at iteration $i$:

$$(\omega S_{00} - \varepsilon_i^s)G_{00} = I + \alpha_i G_{2^i,0}$$
$$(\omega S_{00} - \varepsilon_i)G_{2^i,0} = \beta_i G_{0,0} + \alpha_i G_{2^{i+1},0} \qquad (8)$$

with:

$$\alpha_i = \alpha_{i-1}(\omega S_{00} - \varepsilon_{i-1})^{-1}\alpha_{i-1}$$
$$\beta_i = \beta_{i-1}(\omega S_{00} - \varepsilon_{i-1})^{-1}\beta_{i-1}$$
$$\varepsilon_i = \varepsilon_{i-1} + \alpha_{i-1}(\omega S_{00} - \varepsilon_{i-1})^{-1}\beta_{i-1} + \beta_{i-1}(\omega S_{00} - \varepsilon_{i-1})^{-1}\alpha_{i-1}$$
$$\varepsilon_i^s = \varepsilon_{i-1}^s + \alpha_{i-1}(\omega S_{00} - \varepsilon_{i-1})^{-1}\beta_{i-1} \qquad (9)$$

and where $\alpha_0 = \omega S_{01} - H_{01}$, $\beta_0 = (\omega S_{01} - H_{01})^\dagger$, $\varepsilon_0 = H_{00}$ and $\varepsilon_{0s} = H_{00}$. Once $\alpha_i$ and $b_i$ are small enough, the surface Green's function is obtained from $(\omega S_{00} - \varepsilon_i^s)G_{00} \simeq I$. Subsequently, the non-zero block of the self-energy matrix (for the right contact, block indices -1,-1 in Eq. (10) are consistent with our notation for the semi-infinite lead but correspond to the last block in the device) can be found as:

$$\Sigma_{C,-1,-1}^R = \left[(E + i\eta)S_{-1,0} - H_{-1,0}\right]G_{00}\left[(E + i\eta)S_{0,-1} - H_{0,-1}\right] \qquad (10)$$

The self-energy matrix of the left contact can be computed in a similar way. In Fig. 1d, the good agreement between the $I_D(V_G)$ curves for a $WS_2$ transistor (Fig. 1b) computed with our dissipative orthogonal and non-orthogonal DFT-NEGF models, using $H$ coming from QE+Wannier and OPENMX respectively, is shown, validating our non-orthogonal transport model.

### C. DFT- computed electron-phonon scattering coupling

We can use an analytical and a numerical approach to include scattering in our model. The analytical approach is used here. In the analytical approach, we include two major scattering processes, elastic scattering by acoustic phonons, and inelastic scattering by optical phonons, using an analytical expression for self-energies. In this approach, the self-energy term for elastic acoustic phonon scattering can be approximated as [21]:

$$\Sigma_{S,ac}^< = \frac{\Delta_{ac}^2 k_B T}{\rho V v_p^2} G^<(E), \qquad (11)$$

where $\Delta_{ac}$ (eV) is called the effective deformation potential for elastic scattering, $\rho$ is the density per unit volume, $V$ is the atom volume, $v_p$ is the sound velocity. The above expression is valid only when the acoustic phonon branch has a linear dispersion near the $\Gamma$ symmetry point. As later shown for 2D materials, only the in-plane acoustic phonons (TA and LA) have a linear dispersion and the out-of-plane acoustic phonons (ZA) have a parabolic dispersion (See Fig. 4).

The self-energy term for inelastic scattering by optical phonons can be approximated as [21]:

$$\Sigma_{S,op}^< = \frac{\Delta_{op}^2 \hbar^2}{2\rho V E_{ph}}\left(N_q + \tfrac{1}{2} \pm \tfrac{1}{2}\right) G^<(E \pm E_{ph}) \qquad (12)$$

where $\Delta_{op}$ (eV/cm) is the effective deformation potential for optical phonon scattering and $E_{ph}$ is the optical phonon energy. We extract the parameters ($\Delta_{ac}$, $v_p$, $\Delta_{op}$ and $E_{ph}$) by calculating phonon dispersions and electron-phonon matrix elements using the density functional perturbation theory (DFPT) approach [10,25]. Below, considering monolayer $WS_2$ as an example, we describe the procedure of extracting these parameters.

In Fig. 4, we show the phonon dispersion for $WS_2$ calculated using DFPT as implemented in QE [10,25].

Monolayer $WS_2$ has nine phonon modes. The three lower energy branches are acoustic phonons and the remaining six represent the optical phonons. $v_p$ for the in-plane acoustic modes can be extracted by taking the slope of the linear region from the dispersion near the Γ point. For materials like WS$_2$, which have a horizontal mirror symmetry, scattering by ZA phonons are found to be negligible [26]. Therefore, its contribution can be ignored.

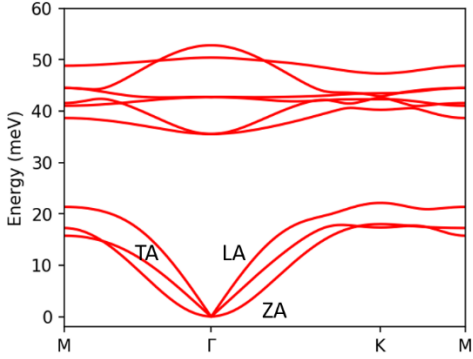

Fig. 4. Phonon dispersion of WS$_2$ calculated using the DFPT approach.

The plane-wave based electron-phonon coupling matrix elements calculated from DFPT are of the form [10,25]:

$$M_{mn\lambda}(\mathbf{k}\eta, \mathbf{k'}\eta') = \left(\frac{\hbar}{2M_{\text{cell}}\omega_\mathbf{q}}\right) DK_\lambda(\mathbf{k}\eta, \mathbf{k'}\eta') \quad (13)$$

where **k** and **k'** are the initial and final electron wave vectors, respectively, η and η' are the initial and final bands, respectively, λ is the phonon mode, **q** = **k'**-**k** is the phonon wave vector, $M_{\text{cell}}$ is the total mass of the unit cell and DK is called the deformation potential.

For small **q** vector, DK and $\Delta_{ac}$ can be related as: $DK \approx \Delta_{ac}|\mathbf{q}|$. Therefore, to extract $\Delta_{ac}$, we calculate DK from the matrix elements, for the same initial and final electron energies, as functions of the angle of the final wave vector formed with respect to the direction of the initial wave vector. The initial wave vector **k** is taken along the Γ-K direction. $\Delta_{ac}$ is then obtained by dividing DK by the magnitude of the phonon wave vector |**q**|. In Fig 5, we show DK and $\Delta_{ac}$ plotted for different initial and final electron kinetic energies for LA phonons. In the case of monolayer WS$_2$ we find that $\Delta_{ac}$ does not vary significantly for different kinetic energies or scattering angle. Therefore, it can be approximated as a constant.

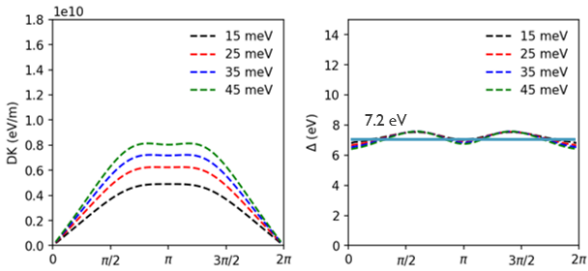

Fig. 5. Acoustic deformation potential (*DK*) and effective deformation potential ($\Delta_{ac}$) plotted as a function of the angle of the final wave vector formed with respect to Γ-K direction.

For optical phonons, $\Delta_{op} = DK$. Therefore, we calculate DK from the matrix elements for different optical modes for an initial wave vector **k** at the conduction band minima and the final wave vector **k'** lying on an equi-energy surface with an energy equal to the optical phonon energy $E_{\text{ph}}$. In Fig. 6, we show the effective optical deformation potentials and their corresponding phonon energies, $E_{\text{ph}}$ plotted as a function of the angle of the final wave vector formed with respect to Γ-K direction. We find that DK is constant for all the optical phonon modes.

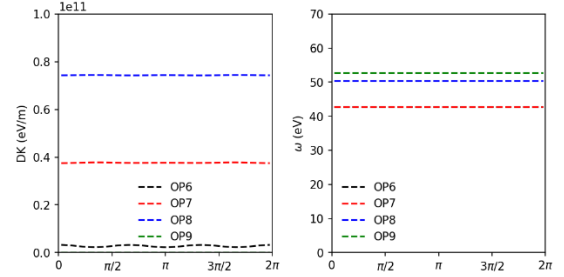

Fig. 6. Optical deformation potential (DK) and optical phonon energy ($E_{\text{ph}}$) plotted as a function of the angle of the final wave vector formed with respect to Γ-K direction.

The analytical approach is computationally efficient and can be a good approximation when we obtain constant effective deformation potentials. However, it has been shown that for 2D materials with an anisotropic band structure, such as phosphorene, $\Delta_{ac}$ has a strong angular dependency and approximating with a constant would lead to an error [27]. In addition, certain 2D materials like silicene and germanene lack horizontal mirror symmetry. Therefore, scattering by ZA phonons is not negligible [26]. For this case, we are currently developing a numerical approach, where we use full electron-phonon matrix elements and phonon dispersion obtained from DFPT to evaluate Eq. (6).

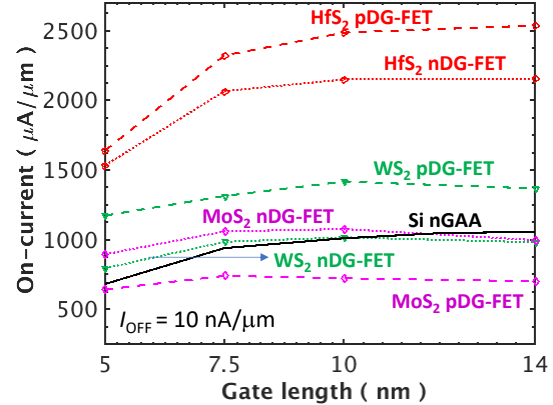

Fig. 7. DFT-NEGF simulated maximum achievable $I_{\text{ON}}$ vs. *L* for optimized n-and p-type TMD DG MOSFETs made of MoS$_2$, WS$_2$ and HfS$_2$. The NEGF simulated achievable $I_{\text{ON}}$ of the Si n-type Gate-all-around (GAA) MOSFET is also shown for comparison. EOT = 0.5 nm. $V_{DD}$ = 0.6 V. $I_{\text{OFF}}$ = 10 nA/mm. The current is normalized by the gate perimeter. Electron-phonon scattering is included.

### III. APPLICATIONS

*A. Novel material and device exploration for ultra-scaled CMOS.*

As an application, we have used our simulator to explore the fundamental physics and performance of 2D materials for sub-10 nm CMOS [6]. Figure 7 compares the DFT-NEGF simulated maximum achievable on-current, $I_{\text{ON}}$, vs. *L* at a fixed typical high-performance (HP) off-state leakage $I_{\text{OFF}}$ =10 nA/μm for optimized n-and p-type TMD Double-gated (DG) MOSFETs (Fig. 1b) made of monolayer (1ML) MoS2, WS$_2$ and HfS$_2$. The simulated achievable $I_{ON}$ of the Si n-type Gate-all-around (GAA) MOSFET is also shown for comparison. The $HfS_2$ transistors feature a promising boost of on-current vs. Si at all gate lengths. $WS_2$ CMOS could also be interesting owing to its strong p-type transistor drive.

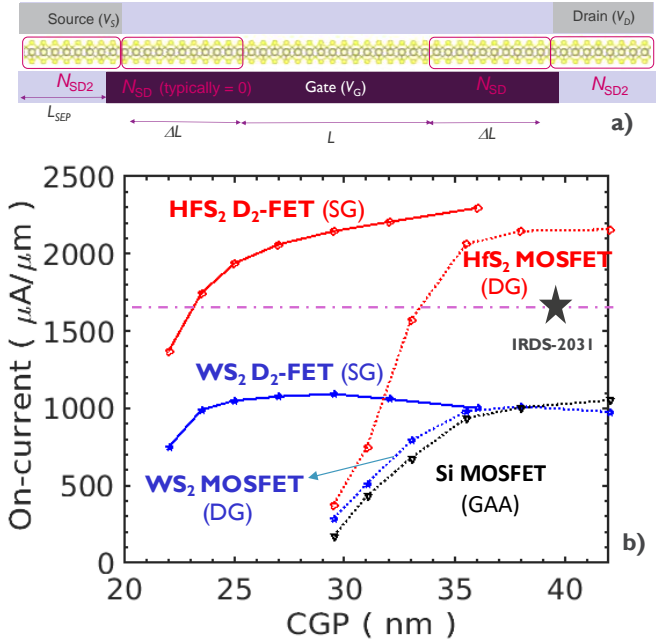

Fig. 8. a) Schematic of the single-gated (SG) $D_2$-FET device. b) DFT-NEGF simulated $I_{ON}$ vs. CGP for optimized n-type MOSFETs and $D_2$-FETs made of $WS_2$ and $HfS_2$. A DG architecture is assumed for all the 2D-MOSFETs, while a SG-architecture is employed for the 2D $D_2$-FETs. The Si GAA n-MOSFET $I_{ON}$ vs. CGP is also shown for comparison. EOT = 0.5 nm. $V_{DD}$ = 0.6 V. $I_{OFF}$ = 10 nA/μm. The current is normalized by the gate perimeter. For the $D_2$-FETs, we used $\Delta L = L/2$ with a minimum value of $\Delta L$ = 4 nm for $L \leq 8$ nm.

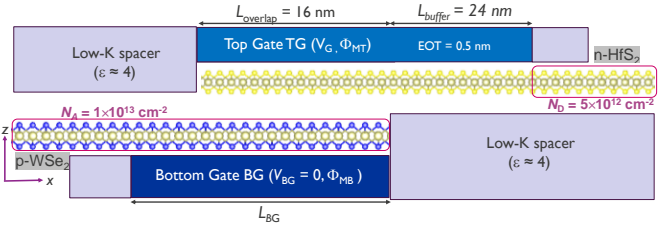

Fig. 9. Schematic view of the optimized $HfS_2/WSe_2$ nTFET.

Even using DG 1ML 2D-material MOSFETs, it is not possible to keep the performance while down scaling $L$ below 5 nm [6], that is, scaling the contacted gate pitch, CGP (i.e., the minimal distance between the gate of 2 subsequent transistors) below 33 nm. CGP is composed of the sum of $L + 2 \times L_{SD}$. $L_{SD}$ = 14 nm is the length of the highly-doped source-and-drain extensions (see Fig. 1b). This $L_{SD}$ value is obtained using the 2031 IRDS-dimensional targets for the so-called 1-nm-technology node and beyond [28]. By using a dynamically-doped transistor, $D_2$-FET, design (Fig. 8) that scales better than the MOSFET [6], we can further scale CGP. Combining a high-mobility 1ML-2D channel material such as $HfS_2$ with the $D_2$-FET architecture, our results predict that the stringent 2031 IRDS HP current target, that was derived for a pitch of 40 nm, can be achieved but with a pitch of about 22 nm instead and using a single extended back gate only (Fig. 8). The $D_2$-FET success for scaling comes from the individually-gated back-gate concept that does not require spacers with the top metal contacts and allows for dynamically doping what used to be the source-and-drain extensions ($N_{SD}$-doped region in Fig. 8a) in the MOSFET design [6]. In the $D_2$-FET, $N_{SD}$ is typically left intrinsic and doping under the metal contact ($N_{SD2}$) is only required if a Schottky barrier is present at the interface.

### B. $HfS_2/WSe_2$ VDW TFET with Wannierized H

The tunneling current in a vdW TFET has been shown to strongly increase with the interlayer coupling [3]. In order to boost the drive-current, a material system that would achieve a broken-gap and maximize the interlayer coupling would be desired. Here, we assess the potential of a $HfS_2/WSe_2$ vdW TFET. This specific choice of materials results from a preliminary DFT study of the charge transfer with respect to a perpendicularly applied electric field. The study was done for over about 140 pairs of heterostacked vdW materials [4]. A stronger charge transfer between the layer may be related to a stronger interlayer coupling, hence favor tunneling. The $HfS_2/WSe_2$ stack was found to have the highest charge transfer, as well as a broken gap. $HfS_2$ and $WSe_2$ are promising material for n- and p-MOS transistors respectively [6].

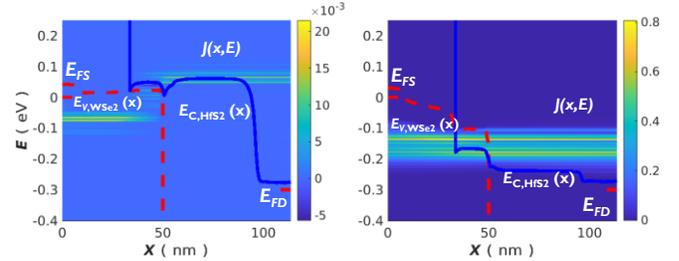

Fig. 10. Current spectrum $J(E,x)$ (surface plot), as well as top conduction-band ($E_{C,HfS2}$) (-) and bottom valence band ($E_{V,WSe2}$) (--) edges along the channel direction, $x$, of the $HfS_2/WSe_2$ nTFET with e-ph of Fig. 8 a) in off- and b) on-state.

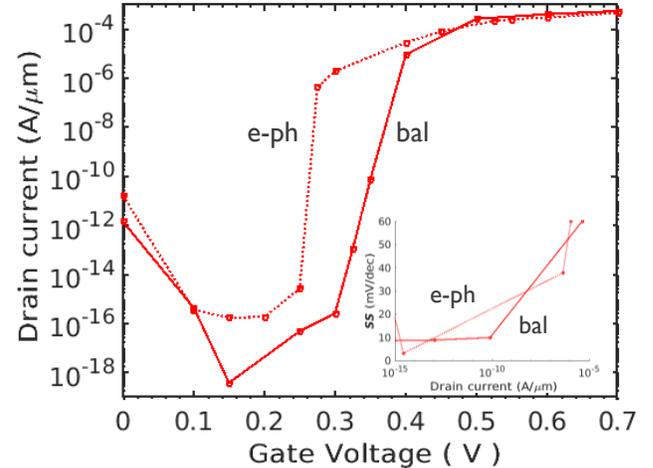

Fig. 11. Ballistic (bal) and dissipative (e-ph) $I_D(V_G)$ and SS($I_D$) (inset) characteristics of the optimized $HfS_2/WSe_2$ vdW nTFETs. $V_{DD}$ = 0.3 V.

The schematic of the optimized nTFET design is presented in Fig. 9. The device transport axis is chosen aligned with the ΓM orientation (indirect band gap) to maximize the ballistic current [2]. This vdW TFET is a line TFET. Its design and optimization are similar to the Core-Shell InAs/GaSb TFET for which a detailed study was presented [29]. The basic functioning in off- and on-state is illustrated in Fig. 10, showing band diagrams and current spectrum flows in the device. The bottom $WSe_2$ layer acts as the source and is assumed p-doped with a density of $1 \times 10^{13}$ cm$^{-2}$. The device switching is controlled by the top gate (TG). TG extends over the 16 nm region where top and bottom layers overlap and for another 24 nm toward the drain extension. In off-state (Fig. 10a), the top conduction-band, $E_{C,HfS2}$, and bottom valence band, $E_{V,WSe2}$, are not overlapping and the BTBT current is strongly reduced. In on-state (Fig. 10b), the TG electrostatic potential pushes $E_{C,HfS2}$ to a lower energy than $E_{V,WSe2}$, and a high BTBT current flows through the device. The 40 nm-long $HfS_2$ top-layer part of the device under TG (16 nm channel and 24 nm buffer layer) is assumed intrinsic. The drain part is n-doped to $0.6 \times 10^{13}$ cm$^{-2}$. The role of the buffer layer is to prevent off-state point tunneling leakage from the bottom layer to the drain, similarly to what was show in [29].

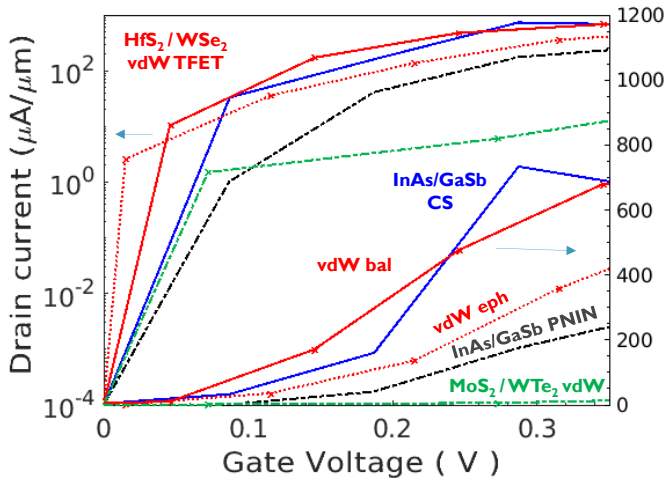

Fig. 12. Ballistic (bal) and dissipative (e-ph) $I_D(V_G)$ characteristics of the optimized $HfS_2/WSe_2$ vdW-, and InAs/GaSb PNIN [30, 11], and CS [29] GAA NW nTFETs. The $I_D(V_G)$ characteristics (with e-ph) of the optimized $MoS_2/WTe_2$ vdW TFET are also shown. $V_{DD}$ = 0.35 V. $I_{OFF}$ = 100 pA/µm.

To prevent from unwanted gating and depletion by TG, a grounded bottom gate (BG) with a higher work function ($\Phi_{MB} - \Phi_{MT}$ = 0.7 eV) induces an additional p-type carrier concentration in the bottom-layer. BG extends under the 16 nm overlap region and for several nm toward the source-side. The low-K dielectric spacer over the source, with a relative permittivity close to 4, further reduces the TG fringing field that is detrimental to on-current. One unwanted consequence of the different material and charge environment (e.g., a high bottom-layer charge under the overlap region) is that the channel top-layer (conduction) bands may be shifted upward in energy compared to those in the buffer layer. Tuning $\Phi_{MB}$ further helps re-aligning the bands. This prevents unwanted point-edge tunneling from the edge of the bottom layer to occurs before the full turning on of the device [29].

The impact of electron-phonon (e-ph) scattering on the $I_D(V_G)$ characteristics of the vdW TFET is shown in Fig. 11. Compared to the ballistic case, an earlier onset of conduction and a reduction of the minimum leakage floor is observed as collisions with inelastic phonons allow for the BTBT electronic current to flow although the tunneling window is not yet fully opened (Fig. 10a). Both characteristics, however, show abrupt switching ($S$ < 10 mV/dec) with potential for high $I_{ON}$ of 680 and 420 µA/µm at $V_{DD}$ = 0.35V and $I_{OFF}$ = 100 pA/µm in the ballistic and e-ph cases respectively (Fig. 12). Even in the more realistic case, when e-ph is included, the achieved drive current, although lower than that of the III-V CS-TFET [29], is about twice the drive current delivered by an optimized InAs/GaSb PNIN GAA TFET [30] and between 3 to 4 times higher than the DFT-NEGF-simulated ballistic current predicted for the $MoTe_2/SnS_2$ vdW TFET [2]. This high level of current is in line with our expectations of a high interlayer-coupling coefficient favoring BTBT current for this specific vdW stack.

In Fig. 12, we also present the DFT-NEGF simulated $I_D(V_G)$ characteristics of the $MoS_2/WTe_2$ vdW TFET. Our computations predict an interesting 10 meV staggered direct gap at the K-point for this material combination. These findings are in line with results in [7] and [8]. Our DFT-NEGF predicted $I_{ON}$, however, is a modest 14 µA/µm, that we attribute to a low vdW interlayer coupling. In [7] and [8] the modeled drive current was equal or greater than 1000 µA/µm, probably indicating that a very high interlayer coupling was assumed. Such level of current is equal or greater than that we get with the $MoS_2$ MOSFET at $V_{DD}$ = 0.6V. [7] and [8] used non-ab-initio models fitted on DFT band structures. Such models tend to strongly overestimate drive current for MOSFETs [6]. An extra complication arises using these models for vdW TFETs, as one has to estimate/model the small vdW interlayer coupling, which is difficult in practice.

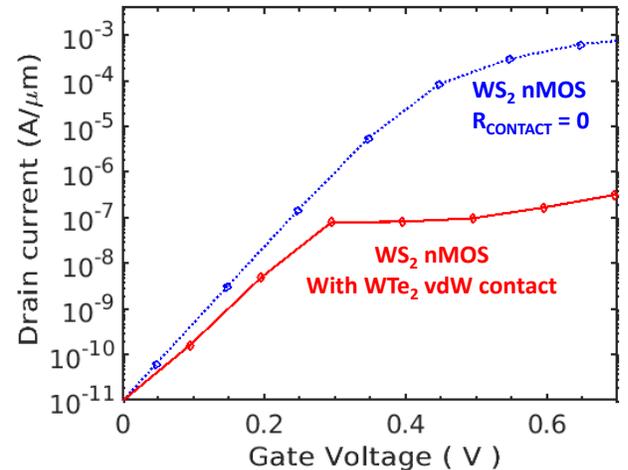

Fig. 13. $I_D(V_G)$ characteristics of a $L$ = 14 nm $WS_2$ nFET with (Fig. 3b) and without (Fig. 1b) a $WTe_2$ vdW top contact using the Non-orthogonal NEGF model and the OPENMX $H$ described in section II. $V_{DD}$ = 0.6 V. e-ph is included.

### C. $WTe_2$ / $WS_2$ VDW Metal / semiconductor non-orthogonal transport.

Finding a metal with a low Schottky barrier to achieve a low contact resistance is one of the key challenges to address towards 2D-material CMOS. 2D-metal/2D semiconductor vdW contacts have been shown to be an interesting option, as they may be free of Fermi-level pinning [5]. Our non-orthogonal DFT method is well-suited to screen such material combination, especially if the supercells become large. As an example, in Fig. 12, we show the $I_D(V_G)$ characteristics of the simulated $WS_2$ device (Fig. 3b) using the WTe$_2$/WS$_2$ heterostack contact of Fig. 3a. Due to the rather high SB$_H$ of about 0.5 eV (Fig. 3c) and the vdW junction, the current is strongly limited in on-state compared to a WS$_2$ transistor that assume an ideal ohmic contact.

## IV. Conclusion

We have presented, here, advanced DFT-NEGF techniques to explore transport in novel materials and devices and in particular in vdW heterojunction transistors. Aside from the popular method using plane-wave DFT, followed by a Wannierization step, we have detailed our LCAO-DFT approach. This technique is more efficient when supercells become larger but results in a non-orthogonal NEGF model. Such model was described here, including the Sancho-Rubio and electron-phonon scattering within a non-orthogonal framework. We also presented our methodology to extract electron-phonon coupling from first principle and to include it in our transport simulations using analytical isotropic deformation potentials. We are currently working towards the implementation of a numerical method that directly use the non-isotropic full matrix and that can be useful for non-isotropic materials. Finally, we apply our methods towards the exploration of novel 2D materials and devices. This includes 2D material selection and the Dynamically-Doped FET for ultimately scaled CMOS, the exploration of vdW TFETs, in particular the HfS$_2$/WSe$_2$ TFET that could achieve high on-current levels, and the study of Schottky-barrier height and transport through a metal-semiconducting WTe$_2$/WS$_2$ vdW junction transistor.

## References

[1] T. Roy, M. Tosun, J. S. Kang, A. B. Sachid, S. B. Desai, M. Hettick, C. C. Hu, and A. Javey, "Field-Effect Transistors Built from All Two-Dimensional Material Components", ACS Nano, vol. 8, 6259 (2014).

[2] A. Szabó, S. J. Koester, and M. Luisier, "Ab-Initio Simulation of van der Waals MoTe$_2$–SnS$_2$ Heterotunneling FETs for Low-Power Electronics", IEEE Elec. Dev. Letters, vol. 36, no. 5, pp. 514-516, 2015. https://doi.org/ 10.1109/LED.2015.2409212.

[3] K.-T. Lam, G. Seol and J. Guo, "Performance Evaluation of MoS2-WTe2 Vertical Tunneling Transistor using Real-space Quantum Simulator", in Proc. IEEE Int. Electron Device Meeting (IEDM), San-Francisco, CA, USA, 2014, pp. 30.3.1-30.3.4. https://doi.org/10.1109/IEDM.2014. 7047141.

[4] A. K. A. Lu, M. Houssa, I. P. Radu, and G. Pourtois, ACS Appl. Mater. Interfaces, 2017, 9 (8), pp. 7725–7734.

[5] S. Song *et al.*, "Wafer-scale production of patterned transition metal ditelluride layers for two-dimensional metal–semiconductor contacts at the Schottky–Mott limit", Nat Electron 3, 207–215 (2020). https://doi.org/10.1038/s41928-020-0396-x

[6] A. Afzalian, "Ab initio perspective of ultra-scaled CMOS from 2D-material fundamentals to dynamically doped transistors", npj 2D Mater Appl 5, 5 (2021). https://doi.org/10.1038/s41699-020-00181-1

[7] J. Cao *et al.*, "Operation and Design of van der Waals Tunnel Transistors: A 3-D Quantum Transport Study," in IEEE Transactions on Electron Devices, vol. 63, no. 11, pp. 4388-4394, Nov. 2016, doi: 10.1109/TED.2016.2605144.

[8] F. Chen et al., "Transport in vertically stacked hetero-structures from 2D materials", Journal of Physics: Conference Series, vol 864, p. 012053, Jun. 2017. https://doi.org/10.1088/1742-6596/864/1/012053

[9] G. Kresse and J. Furthmuller, "Efficient iterative schemes for ab initio total-energy calculations using a plane-wave basis set", Phys. Rev. B, vol. 54, no. 11, pp. 11169-11186, 1996. https://doi.org/10.1103/PhysRevB.54.11169.

[10] P. Giannozzi *et al.*, "QUANTUM ESPRESSO: a modular and open-source software project for quantum simulations of materials". J. Phys. Condens. Matter., vol. 21, no. 39, p. 395502, 2009. https://doi.org/10.1088/0953-8984/21/39/395502.

[11] A. Afzalian *et al.*, "Physics and performance of III-V nanowire broken-gap heterojunction TFETs using an efficient tight-binding mode-space NEGF model enabling million-atom nanowire simulations", J. Phys. Condens. Matter, vol. 30, no. 25, 254002 (16pp), 2018. https://doi.org/10.1088/1361-648X/aac156.

[12] A. Calzolari, N. Marzari, I. Souza and M. Buongiorno Nardelli, "Ab initio transport properties of nanostructures from maximally localized Wannier functions", Phys. Rev. B, vol. **69**, p. 035108, 2004.

[13] A. Afzalian and G. Pourtois, "ATOMOS: An ATomistic MOdelling Solver for dissipative DFT transport in ultra-scaled HfS2 and Black phosphorus MOSFETs", 2019 International Conference on Simulation of Semiconductor Processes and Devices (SISPAD), Udine, Italy, 4-6 Sept. 2019. DOI: 10.1109/SISPAD.2019.8870436.

[14] M. C. Neale *et al.*, "OpenMx 2.0: Extended structural equation and statistical modeling." Psychometrika, vol. **81**, no. 2, pp. 535-549, 2016. doi: 10.1007/s11336-014-9435-8.

[15] T. D. Kuhne *et al.*, "CP2K: An electronic structure and molecular dynamics software package - Quickstep: Efficient and accurate electronic structure calculations", The Journal of Chemical Physics, vol. 152, no. 19, p. 194103, 2020, https://doi.org/10.1063/5.0007045.

[16] S. Baroni, S. de Gironcoli, A. Dal Corso, P. Giannozzi, "Phonons and related crystal properties from density-functional perturbation theory", Rev. Mod. Phys., vol. 73, no. 2, pp. 515-562, 2001. https://link.aps.org/doi/10.1103/RevModPhys.73.515.

[17] A. Laturia, M.L. Van de Put and W.G. Vandenberghe, "Dielectric properties of hexagonal boron nitride and transition metal dichalcogenides: from monolayer to bulk", npj 2D Mater Appl, vol. 2, no. 6, 2018.

[18] J. Klimeš, D. R. Bowler and A. Michaelides, "Chemical accuracy for the van der Waals density functional", J. Phys.: Cond. Matt., vol. 22, p. 022201, 2010.

[19] S. Grimme, J. Antony, S. Ehrlich and H. Krieg, "A consistent and accurate ab initio parametrization of density functional dispersion correction (DFT-D) for the 94 elements H-Pu", J. Chem. Phys, vol. 132, no. 15, pp. 154104, 2010. Doi: 10.1002/jcc.20495.

[20] A.A. Mostofi, J.R. Yates, G. Pizzi, Y.S. Lee, I. Souza, D. Vanderbilt, N. Marzari, "An updated version of wannier90: A tool for obtaining maximally-localised Wannier functions", Comput. Phys. Commun. 185, 2309 (2014).

[21] A. Afzalian, "Computationally Efficient self-consistent Born approximation treatments of phonon scattering for Coupled-Mode Space Non-Equilibrium Green's Functions", J. Appl. Phys., vol. 110, p. 094517, 2011. https://doi.org/10.1063/1.3658809.

[22] M. Shin, et al., "Density functional theory based simulations of silicon nanowire field effect transistors," J. Appl. Phys., vol. 119, no. 15, p. 154505, Apr. 2016.

[23] Y. Wu, et al. "Conductance of graphene nanoribbon junctions and the tight binding model," Nanoscale Res. Lett., vol. 6, no. 1, p. 62, Dec. 2011.

[24] J.-J. Zhou et al., "Perturbo: a software package for ab initio electron-phonon interactions, charge transport and ultrafast dynamics." ArXivpreprint arXiv:2002.02045 (2020).

[25] M. V. Fischetti, and W. G. Vandenberghe. "Mermin-wagner theorem, flexural modes, and degraded carrier mobility in two-dimensional crystals with broken horizontal mirror symmetry." Physical Review B 93.15 (2016): 155413.

[26] G. Gaddemane et al., "Theoretical studies of electronic transport in monolayer and bilayer phosphorene: A critical overview." Physical Review B 98.11 (2018): 115416.

[27] https://irds.ieee.org/editions/2018.

[28] A. Afzalian, G. Doornbos, T.-M. Shen, M. Passlack and J. Wu, "A High-Performance InAs/GaSb Core-Shell Nanowire Line-Tunneling TFET: An Atomistic Mode-Space NEGF Study", IEEE J. of Electron Dev. Society, Nov. 2018, DOI: 10.1109/JEDS.2018.2881335.

[29] A. Afzalian, M. Passlack and Y.-C. Yeo, "Scaling perspective for III-V broken gap nanowire TFETs: An atomistic study using a fast tight-binding mode-space NEGF model", in Proc. IEEE Int. Electron Device Meeting (IEDM), San-Francisco, CA, USA, 2016, pp. 30.1.1-30.1.4. https://doi.org/10.1109/IEDM.2016.7838510.